\pdfoutput=1

\documentclass[aps,preprint,groupedaddress]{revtex4-1} 
\usepackage{amsmath}
\usepackage{graphicx}

\usepackage{color}

\def\bra#1{\mathinner{\langle{#1}|}}
\def\ket#1{\mathinner{|{#1}\rangle}}
\newcommand{\braket}[2]{\langle #1|#2\rangle}
\newcommand{\ketbra}[2]{|#1\rangle\langle#2|}

\newcommand{\up}{\uparrow}
\newcommand{\dn}{\downarrow}
\newcommand{\B}[1]{\mathbf{#1}}

\begin{document}

\title{Control of inhomogeneous atomic ensembles of hyperfine qudits}


\author{Brian E. Mischuck} \email{bmischuck@gmail.com}
\affiliation{Lundbeck Foundation Theoretical Center for Quantum System Research,}
\affiliation{Department of Physics and Astronomy, Aarhus University, DK-8000 Aarhus C, Denmark}
\author{Seth T. Merkel} 
\affiliation{Institute for Quantum Computing}
\affiliation{Department of Physics and Astronomy, University of Waterloo, Waterloo, ON N2L 3G1, Canada}
\affiliation{Theoretical Physics, Saarland University, D-66041 Saarbr\"{u}cken, Germany}
\author{Ivan H. Deutsch}
\affiliation{Center for Quantum Information and Control (CQuIC)}
\affiliation{Department of Physics and Astronomy, University of New Mexico, Albuquerque NM 87131}

\date{\today}

\begin{abstract}
We study the ability to control $d$-dimensional quantum systems (qudits) encoded in the hyperfine spin of alkali-metal atoms through the application of radio- and microwave-frequency magnetic fields in the presence of inhomogeneities in amplitude and detuning.  Such a capability is essential to the design of robust pulses that mitigate the effects of experimental uncertainty and also for application to tomographic addressing of particular members of an extended ensemble. We study the problem of preparing an arbitrary state in the Hilbert space from an initial fiducial state.  We prove that inhomogeneous control of qudit ensembles is possible based on a semi-analytic protocol that synthesizes the target through a sequence of alternating rf and microwave-driven $SU(2)$ rotations in overlapping irreducible subspaces. Several examples of robust control are studied, and the semi-analytic protocol is compared to a brute force, full numerical search.  For small inhomogeneities, $< 1\%$,  both approaches achieve average fidelities greater than 0.99, but the brute force approach performs superiorly, reaching high fidelities in shorter times and capable of handling inhomogeneities well beyond experimental uncertainty.  The full numerical search is also applied to tomographic addressing whereby two different nonclassical states of the spin are produced in two halves of the ensemble.

\end{abstract}


\maketitle

\section{\label{intro}Introduction}

Motivated by applications ranging from controlling chemical dynamics to quantum information processing (QIP), control over quantum systems has become an increasingly important tool \cite{nielsen_quantum_2000, boixo_generalized_2007, rabitz_chemistry:_2003, judson_teaching_1992, khaneja_time_2001, merkel_constructing_2009}. In the context of QIP, the majority of studies of such processes are based on a collection of two-level ($d=2$) subsystems or qubits, a natural extension of binary classical logic.  In practice, the physical objects that encode quantum information never have solely two levels and control is necessary to isolate a particular qubit of interest.  Alternatively,  higher dimensional $d>2$ systems, or qudits, can be employed for base-$d$ quantum logic.  While from a computer science perspective encoding in qudits does not lead to a change in computation complexity, from a physical perspective, information processing with qudits provides different trade-offs. For example, a $\log_2 d$ reduction in the number of subsystems must be compared with the complexity in implementing the universal $SU(d)$ gates and quantum error correcting codes \cite{grassl_2003}.  Moreover, qudits are interesting in their own right exhibiting properties such as nonlocality without entanglement \cite{bennet_1999} and providing a platform for explorations of quantum chaos  \cite{chaudhury_quantum_2009}.  

A natural qudit is the hyperfine manifold of magnetic sublevels associated with the ground-electronic state of atoms, providing a Hilbert space of dimension $d=(2J+1)(2I+1)$ where $J$ is the electron angular momentum and $I$ is the nuclear spin.  Motivated by ongoing experiments in the Jessen group \cite{chaudhury_quantum_2007}, we work with $^{133}$Cs, an alkali-metal atom with one valence electron $J=S=1/2$ and nuclear spin $I=7/2$, yielding hyperfine coupled spins of magnitude $F=I \pm S=3,4$, and a total Hilbert space of dimension $d=16$.   Extensions to other elements, such as the rare-earths, open opportunities to even larger Hilbert spaces (e.g., and $d=128$ in holmium \cite{saffman_2008}).  Early proposals for qudit control in hyperfine manifolds involved the use of a series of two-photon Raman transitions between magnetic sublevels \cite{oleary_parallelism_2006}.  More recently, time-dependent magnetic fields and a static tensor light shift have been used for arbitrary state preparation in the lower hyperfine manifold \cite{chaudhury_quantum_2007}.  Alternative schemes that employ radio-frequency and microwave magnetic fields have also been studied for applications in qudit quantum control \cite{merkel_quantum_2008, merkel_constructing_2009, riofrio_2011}.  

Our goal in this article is to expand the control tool box for hyperfine qudits, with attention to alkali-metal atoms, the standard elements used in laser cooling experiments.   We particularly develop the methods of {\em ensemble control}, previously studied for spin-1/2 nuclei in the context of NMR \cite{khaneja_optimal_2005}, whereby subsystems are subjected to an ensemble of different Hamiltonians.  Such methods are important in a number of scenarios. In the context of experimental uncertainty in the control parameters, the different dynamics generated by different members of the ensemble correspond to errors.  Making the desired dynamics insensitive to this uncertainty is known as ``robust control," and will be essential for the levels of precision required in implementing QIP protocols.  In addition, an inhomogeneous ensemble may be intentionally imposed.  In such a scenario different members of an ensemble may be {\em tomographically addressed} in space or time. 

The idea of robust control goes back to the development of composite pulse techniques for spin-1/2 nuclei in NMR \cite{cummins_tackling_2003, vandersypen_nmr_2005} where special sequences such as ``CORPSE" and ``SCROFULOUS" were designed to perform particular $SU(2)$ rotations in a manner that is robust to errors in detuning and/or Rabi frequency.  In more recent studies, Khaneja and Glaser showed that one can achieve control of spin-1/2 particles as a nearly arbitrary function of detuning or Rabi frequency \cite{li_control_2006, kobzar_pattern_2005, khaneja_optimal_2005}.  Researchers are also applying such NMR-inspired tools to the control of cold atoms.  For example, the groups of Blatt and Chuang used robust pulses to improve the performance of quantum algorithms with trapped ions \cite{gulde_2003}, and Jessen's group explored robust control neutral atom qubits trapped in optical lattices \cite{rakreungdet_accurate_2009}. Tomographic addressing of ultracold atoms is becoming increasingly important.  Using real magnetic fields, the Bloch group addressed the ``shells" of a Mott insulator of \cite{flling_formation_2006} and the Meschede group addressed sites of a one dimension lattice lattice beyond the diffraction limit \cite{karski_imprinting_2009}.  Even higher resolution addressing is possible using the spin-dependent light shifts of off-resonant laser fields \cite{deutsch_quantum_2009}.  Using the light shift of a focused laser beam to create a large spatial gradient, Bloch's group individually addressed atoms in a 2D lattice with a spacing closer than the diffraction limit of the addressing beam \cite{weitenberg_2011}.  Such tomographic addressing can benefit from more sophisticated control analysis.

In this article we study robust control and tomographic addressing of hyperfine qudits within the unified framework of ensemble control.  In Sec.~\ref{controls} we describe the physical system and the available control Hamiltonians, and use these  in Sec.~\ref{SA_syn} as the basis for a semi-analytic protocol for synthesizing arbitrary states through a series of $SU(2)$ rotations on overlapping subspaces.  Based on the known results in inhomogeneous control of $SU(2)$ rotations,  we extend our qudit state preparation routine to the case of ensembles.  With the semi-analytics in hand, in Sec.~\ref{sec:num_opt} we present an alternative approach to state synthesis based on numerical optimization.   We compare the two approaches by exploring robust state preparation in the presence of inhomogeneities.  Beyond robust control, we study how we can employ the tools of ensemble control to spatially address different regions of a cold atomic cloud and perform local state preparation to create highly nonclassical, nonequilibrium states of the gas.   In Sec.~\ref{sec:summary} we summarize and present the outlook for future research.

\section{Control Hamiltonian \label{controls}}
  
\begin{figure}
\includegraphics{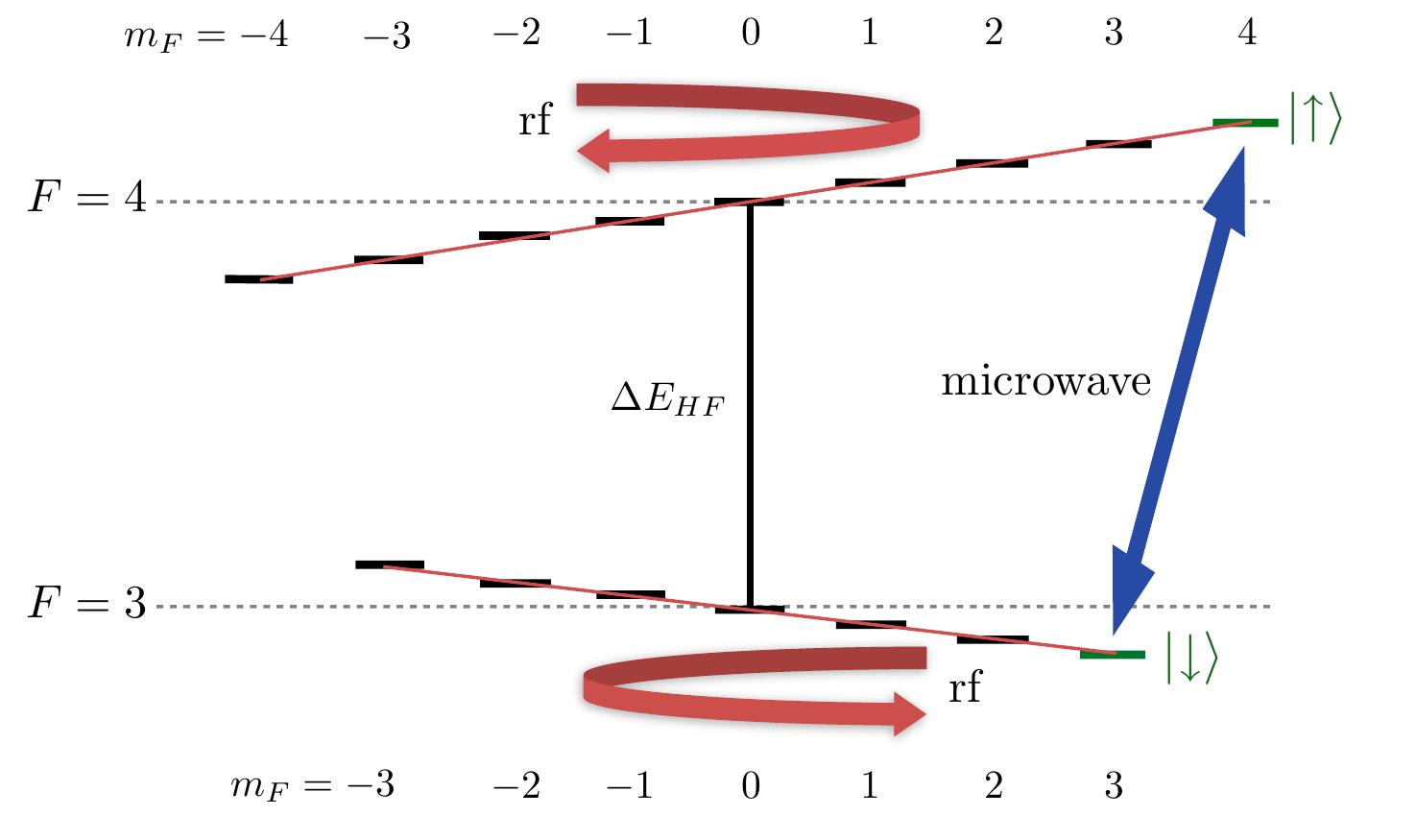}
\caption[rf and Microwave controls]{Radio-frequency (rf) magnetic fields  (in red) and microwave magnetic fields (in blue) control the ground-electronic hyperfine manifold of magnetic sublevels in $^{133}$Cs.  $SU(2)$ rotations in  the irreducible subspaces, $F=3,4$, are driven by rf-magnetic fields.  Because of the sign difference of the $g$-factors for the upper and lower manifold, the two rotations are in opposite directions.  The microwave field is shown to be resonant only with the transition  between the states $\ket{F=4,m_F=4}$ and $\ket{F=3,m_F=3}$  due to the application of a bias magnetic field that breaks the degeneracy between the magnetic sublevels. Microwave-driven $SU(2)$ rotations on this qubit, together with the rf-control, can generate an arbitrary unitary transformation on the full 16-dimensional manifold.} 
\label{rf_uw_control}
\end{figure}

We will focus on control of the spin state of an alkali-metal atom in its electronic ground state through magnetic interactions.  The governing Hamiltonian of the system is
\begin{equation}
H = A \B{I}\cdot \B{S} +\left( g_s\mu_B \B{S}-  g_I\mu_N  \B{I} \right) \cdot \B{B}(t),
\end{equation}
where the first term represents the hyperfine coupling between the atom's nuclear spin, $\B{I}$, and the valence electron's spin, $\B{S}$, and the second represents the interaction of the spins with the controlling magnetic fields. We consider three contributions to the fields, 
\begin{equation}
\B{B}(t) = B_0\B{e}_z+\B{B}_{rf}(t) + \B{B}_{\mu w}(t),
\end{equation}
whose effect on the energy levels is depicted in Fig.~(\ref{rf_uw_control}).  The first contribution is a static bias magnetic field which splits the energies of the magnetic sublevels in the linear Zeeman regime, $\mu_B B_0 \ll A$, while the next two terms are control fields oscillating at radio- (rf) and microwave-frequencies that drive transitions between those levels.   We work in a regime in which the hyperfine coupling is significantly stronger than the interaction due to the applied magnetic fields.  Then it is convenient to split the state space into a direct sum of spaces with total angular momentum $f^{(\pm)} = I\pm\frac{1}{2}$.  In the linear Zeeman regime, we can use the Land\'e projection theorem to write its contribution as
\begin{equation}
H_{B_0} = \sum_{f=+,-} g_f \mu_B \B{B_0}\cdot \B{F}^{(f)}.
\end{equation}
Defining the Zeeman frequency $\Omega_0 = -g_-\mu_B B_0$ (here and throughout $\hbar=1$), the total static Hamiltonian becomes 
\begin{equation}
H_0 = \frac{\Delta E_{HF}}{2}(P^{(+)}-P^{(-)})+ \Omega_0 (g_r F_z^{(+)}-F_z^{(-)}),
\end{equation}
where $\Delta E_{HF}$ is the hyperfine splitting, and $P^{(\pm)}$ are the projectors on the hyperfine manifolds $f^{(\pm)}$.  To take into account the small difference in magnitude and opposite signs of the g-factors in the lower and upper manifold  arising from the nuclear magneton, we have defined $g_r=|g_+/g_-|$.  
The rf-field resonantly couples magnetic sublevels within a subspace with a given total angular momentum. Defining the Larmor frequencies $\Omega_{x(y)}= -2g_- \mu_B B_{x(y)}$, we can again use the Land\'e projection theorem to write the rf-Hamiltonian as 
\begin{equation}
H_{rf }= 2\Omega_x \cos(\omega_{rf}t-\phi_x)(g_r F_x^{(+)} - F_x^{(-)}) + 2\Omega_y \cos(\omega_{rf}t-\phi_y)(g_r F_y^{(+)} - F_y^{(-)}).
\end{equation}
Transforming into a rotating frame according to $U_{rf} = e^{-i\omega_{rf}t(F_z^{(+)}-F_z^{(-)})}$, and making the rotating wave approximation, yields
\begin{align}
H_{rf} + H_0 =& g_r\left( \Omega_x\, \cos\phi_x  -\Omega_y\, \sin\phi_y \right) F^{\up}_x  + g_r\left( \Omega_x\, \sin\phi_x  +\Omega_y\, \cos\phi_y \right) F^{(+)}_y\nonumber\\
&-\left( \Omega_x\, \cos\phi_x  +\Omega_y\, \sin\phi_y \right) F^{(-)}_x+\left( \Omega_x\, \sin\phi_x  -\Omega_y\, \cos\phi_y \right) F^{(-)}_y\nonumber\\
&+\Delta(g_r F_z^{(+)}-F_z^{(-)}) +(1-g_r)\omega_{rf}F_z^{(+)}+ \frac{\Delta E_{HF}}{2}(P^{(+)}-P^{(-)}),
\end{align}
where $\Delta = \Omega_0 - \omega_{rf}$.

The microwave power, frequency, and polarization are chosen so as to couple only two magnetic sublevels between the hyperfine manifolds, leaving the other states unaffected due to off-resonance effects.  These states define a pseudo-spin-1/2 $\{\ket{\up}\equiv\ket{f^{(+)},m_\up=f^{(+)}},\ket{\dn}\equiv\ket{f^{(-)},m_\dn=f^{(-)}}\}$ with the usual Pauli operators, and the microwave-driven Hamiltonian takes the standard spin-resonance form,
\begin{equation}
H_{\mu w} = \Omega_{\mu w}\cos(\omega_{\mu w}t-\phi_{\mu w})\sigma_x.
\end{equation}

We now transform into a total rotating frame given by $U_{total} = U_{rf}e^{-i\alpha t(P^{(+)}-P^{(-)})/2}$ with $\alpha = \omega_{\mu w} - (m_\up+m_\dn)\omega_{rf}$ and make the rotating wave approximation.  Choosing the microwave frequency to be resonant with the pseudo-spin transition, $\omega_{\mu w}=(g_+ m_\up-g_- m_\dn)\mu_B B_0+E_{HF}$, the total Hamiltonian in the rotating frame, including the static rf and microwave contributions is 
\begin{align}
H &= g_r\left[ \Omega_x\, \cos\phi_x  -\Omega_y\, \sin\phi_y \right]F^{(+)}_x  + g_r\left[ \Omega_x\, \sin\phi_x  +\Omega_y \,\cos\phi_y \right]F^{(+)}_y\nonumber\\
&-\left( \Omega_x\, \cos\phi_x  +\Omega_y\, \sin\phi_y \right)F^{(-)}_x+\left( \Omega_x\, \sin \phi_x  -\Omega_y\, \cos\phi_y \right)F^{(-)}_y\nonumber\\
&+\frac{1}{2}\Omega_{\mu w} \left( \cos\phi_{\mu w}\, \sigma_x  - \sin\phi_{\mu w}\, \sigma_y  \right)  -\frac{1}{2}(g_r - 1)m_\up\omega_{rf}F^{(+)}_z(P^{(+)}-P^{(-)}).
\end{align}
The control parameters available in this system are the amplitudes and phases of the two rf- and microwave-fields, $\{\Omega_x, \phi_x, \Omega_y, \phi_y, \Omega_{\mu w}, \phi_{\mu w}\}$.  With these controls, we have shown that one can synthesize  any unitary transformation on the hyperfine manifold \cite{merkel_quantum_2008, merkel_constructing_2009}.  
  
If we choose the phases of the $x$ and $y$ rf-coils to be such that $\phi_x = \phi_y-\pi/2 \equiv \phi_{rf}$  and choose the powers in the two coils to be equal, $\Omega_x=\Omega_y=\Omega\equiv \Omega_{rf}$, then the rf polarization is positive-helicity circular and the rf-field is resonant only with the lower hyperfine manifold, leaving the upper manifold fixed.  The Hamiltonian then reduces to 
\begin{subequations}
\begin{align}
H &= H_{rf} + H_{uw}\\
H_{rf} &= 2\Omega_{rf}  \left( -\cos \phi_{rf}\, F^{(-)}_x + \sin\phi_{rf}\, F^{(-)}_y \right), \\
H_{uw}& =  \frac{\Omega_{\mu w}}{2}\left( \cos\phi_{\mu w}\, \sigma_x - \sin\phi_{\mu w}\, \sigma_y \right).
\end{align}
\label{primary_hamiltonian}
\end{subequations}
Through our choice of amplitudes and phases of the applied fields, we have restricted the dynamics to an 8D subspace spanned by the 7D basis in the $F=3$ manifold plus a single auxiliary state in the upper manifold, $\ket{F=4, m_F=4}=\ket{\up}$.  The independent rf or microwave Hamiltonians, $H_{rf}$ and $H_{uw}$, are each generators of an $SU(2)$ rotation in an irreducible subspace.  As we will show in the next section, with the given controls, a unitary transformation on the 8D Hilbert space can be constructed from a sequence of $SU(2)$ rotations.

\section{Control protocols}
\subsection{Semi-Analytic State Synthesis \label{SA_syn}}

We study the problem of synthesizing an arbitrary state within the 8D Hilbert space defined above through a series of $SU(2)$ rotations generated by the control Hamiltonian, Eq.~(\ref{primary_hamiltonian}).  As our fiducial initial state, we begin with all the population in $\ket{\up}$.  As a proof of principle, we employ the technique originally developed by Law and Eberly \cite{law_arbitrary_1996} in which we solve the inverse problem  -- begin with an arbitrary target state $\ket{\psi_T}$ in the space and then map it to the fidulcial state, $\ket{\up}$.  If each of the controls can be reversed, we can then determine the mapping $\ket{\up} \rightarrow \ket{\psi_T}$.  We will see that unlike the original Eberly and Law protocol that involves a qubit, this method will only perform approximate state mapping, with an error that decreases exponentially with the length of the pulse. 

To find a map that transfers the target state  $\ket{\psi_T} \rightarrow \ket{\up}$, we solve a sequence of maximization problems.  First we find an rf-pulse that maximizes the population in the state $\ket{F=3,m_F=3}=\ket{\dn}$ and then we find a microwave-driven rotation that maximizes the population of $\ket{F=4,m_F=4}=\ket{\up}$. The latter transformation can be found analytically because every unitary transformation in this 2$D$ subspace is an $SU(2)$ rotation.   A state 
\begin{equation}
\ket{\chi}=c_{\up}\ket{\up}+c_{\dn}\ket{\dn}
\label{spinor}
\end{equation}
is defined by a Bloch vector, $\B{r}$, for the pseudo-spin.  If $\hat{n}$ is a vector that bisects the $\B{z}$-axis of the Bloch sphere and $\B{r}$,  then a $\pi$-rotation around $\hat{n}$ will drive all the population in $\ket{\chi}$ to $\ket{\up}$. For the rf-driven rotation, the situation is slightly more complex.  The $SU(2)$ matrices represent only a subgroup of the general $SU(2F+1)$ unitary transformations on the space.  Define an $SU(2)$ spin coherent state for $F=3$, parametrized by a direction on the sphere, $\ket{\theta,\phi}\equiv\mathcal{R}^{(3)\dagger}(\theta, \phi)\ket{\dn}$.  We we thus seek the $F=3$ irrep of the $SU(2)$ rotation $\mathcal{R}^{(3)}(\theta, \phi)$ that satisfies
\begin{equation}
\max_{\theta, \phi} |\bra{\dn} \mathcal{R}^{(3)}(\theta, \phi) \ket{\psi_T}|^2.
\label{ideal_husimi}
\end{equation}
This is equivalent to the maximum value of the Husimi distribution with respect to spin-coherent states, $Q(\theta,\phi)= |\braket{\theta,\phi}{\psi_T}|^2$, which we can find easily.

How optimal is this procedure in the achievable fidelity and the required number of steps?  Since the microwave-driven rotation can completely transfer all population from the $\ket{\dn}$ state to $\ket{\up}$,  the only question is how much population in the lower manifold we can transfer to $\ket{\dn}$ using rf-driven rotations.  Because the Husimi distribution $Q(\theta,\phi)$ is everywhere positive, we can find a lower bound by looking at the case where the Husimi distribution is flat, i.e., the maximally mixed state. For a spin-$F$ the Husmi function in such a case would have a uniform height of  $1/(2F+1)$.  Therefore the amount of population remaining in the lower manifold, and thus our error, is bounded from below by  $(2F/(2F+1))^n$ where $n$ is the number of iterations. This implies that the population decreases exponentially with $n$ and we can rapidly map $\ket{\psi_T} \rightarrow \ket{\up}$.  By reversing the sequence of inverse-rotations, we achieve the desired state preparation, $\ket{\up} \rightarrow \ket{\psi_T}$.

Because this state synthesis protocol is based on a series of $SU(2)$ rotations, we can draw on previously known results in $SU(2)$ control to extend our construction to the case of inhomogeneous parameters.  In particular, Kobzar {\em et al.} \cite{kobzar_pattern_2005} used a powerful numerical routine to find pulse sequences that synthesized arbitrary unitary transformations in two-level systems as a function of inhomogeneous controls, and Li and Khaneja explored the theoretical underpinnings \cite{li_control_2006}.  The key result is that for two-level systems driven by canonical Hamiltonians of the form
\begin{equation}
H[\epsilon,\Delta,\Omega(t),\phi(t)] = (1+\epsilon)\Omega(t) \left(\cos\phi(t) \, \sigma_x +\sin\phi(t) \, \sigma_y \right) + \Delta \sigma_z,
\end{equation}
one can synthesize arbitrary $SU(2)$ transformation,  
\begin{equation}
U(\epsilon,\Delta) = \mathcal{T}\left[ \exp \left( -i \int_0^T H[\epsilon,\Delta,\Omega(t'),\phi(t')] dt' \right) \right],
\end{equation} 
as arbitrary functions of $\epsilon>-1$ and $\Delta$.  Here $\Omega(t)$ and $\phi(t)$ is are the amplitude and phase of the drive waveform that act to control the system.  Whereas we allow for inhomogeneity in the amplitude, the phase is assumed to be known and controlled precisely.   This result generalizes trivially for any irreducible representation of $SU(2)$ on a spin of magnitude $F$.

This  construction implies that  in principle one can design robust waveforms so that $U(\epsilon,\Delta)$ is essentially flat, and that one can design tomographic waveforms that address different members of the ensemble for well-chosen values of  $\epsilon$ and/or $\Delta$.  In practice, the controllability of the system will be limited by practical constraints such as bandwidth, slew rates, and time of interaction.  The duration of the waveform is of particular importance given the ultimate constraint of decoherence.  Note that the waveform that generates a desired $U(\epsilon,\Delta)$, is not unique.  Whereas the proof given by Li and Khaneja provides an analytic algorithm to synthesize $U(\epsilon,\Delta)$, in practice, numerical optimization schemes lead to faster pulse sequences while maintaining very high-fidelity.

We study here an extension of the results of ensemble control of $SU(2)$ rotations to our protocol for qudit control.  The evolution of each member of the ensemble is described by the extension of Hamiltonian, Eq. (\ref{primary_hamiltonian}), 
\begin{align}
H(t)  &= -2\Omega_{rf}(t)(1+\epsilon_{rf})\left(-\cos \phi_{rf}(t) \, F^{(-)}_x + \sin \phi_{rf}(t)\, F^{(-)}_y\right)  \nonumber \\
&+\frac{\Omega_{\mu w}(t)}{2}(1+\epsilon_{\mu w})\left( \cos\phi_{\mu w}(t) \, \sigma_x - \sin\phi_{\mu w}(t) \, \sigma_y \right) \nonumber\\
&+ \Delta g_r m_\up \ket{\up}\bra{\up} - \Delta F^{(-)}_z. 
\label{secondary_Hamiltonian}
\end{align}
Our goal is to synthesize different target states as a function of $\epsilon_{\mu w}$, $\epsilon_{rf}$, and $\Delta$.  For each $SU(2)$ rotation we can draw on the results of Li and Khaneja.  We must ensure that the effect of inhomogeneity still allows efficient coherent transfer of probability amplitude from $F=3$ to $\ket{\up}$, and that controllability is respected in the  sequence of rf- and microwave-driven rotations. 

Because our protocol interleaves  pulses acting on different subspaces, we must carefully examine the controllablity with respect to different inhomogeneities.  Consider first the case of ensemble control in the presence of microwave amplitude inhomogeneity alone.  The first rf-pulse must maximize the population in $\ket{\dn}$, averaged over  $\epsilon_{\mu w}$, so Eq.~(\ref{ideal_husimi}) becomes 
\begin{equation}
\max_{\theta, \phi}\sum_{\epsilon_{\mu w}}  |\braket{\theta,\phi}{\psi_T(\epsilon_{\mu w})}|^2.
\end{equation}
Because the rf pulses cannot distinguish between the different members of this particular ensemble of states, this is equivalent to maximizing the population in $\ket{\dn}$ given an initial state, $ \rho_{eff} $, where 
\begin{equation}
\rho_{eff} = N \sum_{\epsilon_{\mu w}}  \ketbra{\psi_T(\epsilon_{\mu w})} {\psi_T(\epsilon_{\mu w})},
\end{equation}
with $N$ as a normalization constant.  In the worst case $\rho_{eff}$ is a completely mixed state.  In this situation the rf pulse accomplishes nothing, and 1/7 of the population remains in $\ket{\dn}$.  The first microwave $\pi$-pulse will transfers all of the population in $\ket{\dn}$ to $\ket{\up}$.  The second rf pulse acts on another incoherent mixture over $\epsilon_{\mu w}$, but not a completely mixed state because we have emptied all of the population in $\ket{\dn}$.  In the worst case, 1/7 of the remaining 6/7 of the population will be transferred by the rf-pulse to $\ket{\dn}$.  The second microwave pulse in the sequence will in general act on an ensemble of spinors of the form in Eq. (\ref{spinor}), $\ket{\chi(\epsilon_{\mu w})}$, because there now exits an ensemble of complex amplitudes in the superposition between $\ket{\dn}$ and $\ket{\up}$.  However, because we can synthesize an arbitrary $SU(2)$ matrix as a function of $\epsilon_{\mu w}$, we can find a transformation that rotates $\ket{\chi(\epsilon_{\mu w})}$ to $\ket{\up}$ for the relevant range of values of $\epsilon_{\mu w}$.  By repeating this procedure, we continually increase the population in $\ket{\up}$ at an exponential rate, even in the presence of microwave power inhomogeneity.

The case of rf inhomogeneities alone, $\ket{\psi_T(\epsilon_{rf})}$, is less favorable.  Based on the results of Li and Khaneja \cite{li_control_2006}, we can find an $SU(2)$ rotation which maximizes the population in $\ket{\dn}$ for all $\epsilon_{rf}$, and the first microwave $\pi$-pulse will transfer this population to $\ket{\up}$.  Whereas the application of the second rf pulse will again maximize the remaining population in $F=3$ to the state $\ket{\dn}$ for all $\epsilon_{rf}$, the second microwave pulse will generally not work as needed.  Because the microwave field cannot distinguish between different spinors in Eq. (\ref{spinor}) as a function of $\epsilon_{rf}$, $\ket{\chi(\epsilon_{rf})}$, the state seen by the microwaves is mixed, $\rho_{eff} =N \sum_{\epsilon_{rf}}  \ketbra{\chi(\epsilon_{rf})}{\chi(\epsilon_{rf})}  $.  In the worst case, this could be either a state whose Bloch vector points along the $z$-axis of the Bloch sphere or is the completely mixed state.  In either case, the microwaves cannot, on average, increase the amount of population in the $\ket{\up}$ state.  Thus, without further care, this procedure cannot synthesis arbitrary states for different $\epsilon_{rf}$. For the particular case that the target state is independent of $\epsilon_{rf}$, we can overcome this difficulty since we can synthesize a target state at the end of each rf pulse that satisfies this property.  By employing such robust control pulses, we ensure a state $\ket{\chi}$ of the pseudospin that can be rotated by microwaves to $\ket{\up}$ at each stage.  But, more general cases are problematic.

Finally, we consider inhomogeneities in the detuning as might arise from an external magnetic field (either noise or intentionally applied).  In this case we will need to consider the role of reversibility in the state synthesis routine with more care.  According to the Law and Eberly protocol, we design a pulse sequence that maps the target $\ket{\psi_T}$ to the fiducial state $\ket{\up}$, according to $\ket{\up}= U_N...U_2 U_1\ket{\psi_T}$.  Then the desired state preparation follows as $\ket{\psi_T}=  U_1^\dagger U_2^\dagger ...U_N^\dagger \ket{\up}$.  The inverse unitary is generated by the Hamiltonian $-H$.   We can achieve this by simply adding $\pi$ to the phase control waveform $\phi(t)$ and also inverting the detuning, $\Delta \rightarrow -\Delta$.  Thus, if we want to synthesize a target $\ket{\psi_T(\Delta)}$ we must use a sequence of rotations that maps $\ket{\psi_T(-\Delta)} \rightarrow \ket{\up}$ to search for the control waveforms. 

In the presence of detuning inhomogeneity, there is an additional source of errors due to phases accumulated in the rotating frame.  For instance, when the rf pulse is driving the lower manifold for time  $t_{rf}$, the population in the $\ket{\up}$ state will acquire a phase $\phi_{\up} = g_r \Delta m_\up t_{rf}$.  We must compensate for this phase in the design of subsequent microwave pulses. This is possible since we can synthesize arbitrary microwave pulses as a function of detuning.  Similarly, we can compensate for $\Delta$-dependent phases accumulated in the $F=3$ manifold while the microwave pulses are applied through the design of subsequent rf-pulses. 

Similar arguments may be made when multiple inhomogeneities are present.  Based on these arguments we find that arbitrary target states of the form $\ket{\psi_T(\epsilon_{\mu w},\Delta )}$ may be synthesized.   We emphasize that although the particular protocol presented here cannot synthesize states that are different for different $\epsilon_{rf}$, this does not imply that such pulse sequences do not exist.  The existence of a semi-analytic protocol capable of synthesizing states which vary with $\epsilon_{rf}$  is a matter of ongoing research. 

\subsection{Fully-numerical state synthesis \label{sec:num_opt} }

Although the results of the previous section indicate that it is possible to perform qudit ensemble control with a sequence of well-designed $SU(2)$ rotations, the duration of the pulse sequence can be unnecessarily long. We therefore also consider fully numerical optimization to search for the desired controls. The semi-analytic protocol lends confidence that such a search routine will yield high fidelity results. For fully numerical searches we allow the microwaves and rf fields to be applied simultaneously and thus the evolution is no longer a simple series of $SU(2)$ rotations on the given subspaces. We have found that we can attain full control of the system using only piecewise constant variations in the phase.  This has the advantage that we may perform the search via unconstrained optimizations that often converge more rapidly than constrained routines.  Searches then return a vector of the sequence of control parameters over $N$ steps, 
\begin{equation}
\mathbf{\Phi} = (\phi_{rf,1},\phi_{\mu w,1},\phi_{rf,2},\phi_{\mu w,2},...,\phi_{rf,N},\phi_{\mu w,N}).
\end{equation}
To perform the numerical search for control pulses, we choose as our objective function the fidelity 
\begin{equation}
\mathcal{F} ( \mathbf{\Phi} ) =  \sum_{\epsilon_{rf},\epsilon_{\mu w},\Delta} |\bra{\psi_T(\epsilon_{\mu w},\Delta ) }U(\epsilon_{rf},\epsilon_{\mu w},\Delta,\mathbf{\Phi})\ket{\up}|^2,
\label{fidelity}
\end{equation} 
where $\ket{ \psi_T(\epsilon_{\mu w},\Delta ) }$ is the target state for a given value of the ensemble. Unconstrained optimization is performed using the Matlab routine $fminunc$.  

\section{Results}
\subsection{Robust control: semi-analytical vs. fully  numerical search}

We present results from several examples of robust control so that the target state is independent of $\epsilon_{\mu w}$, $\epsilon_{rf}$, and $\Delta$.  The Hamiltonian is described by Eq. (\ref{secondary_Hamiltonian}) with $\Omega_0/2\pi = 100$ kHz.  The maximum rf-Larmor  frequency is  $\Omega_{x,y}/2\pi= 1.5$ kHz, and the maximum microwave-Rabi frequency is $\Omega_{\mu w}/2\pi = 3.5$ kHz.  For these parameters, we respect the linear Zeeman regime and can safely ignore off-resonant effects of the microwave radiation, in line with our model of dynamics on an 8D Hilbert space -- the span of the $F=3$ subspace plus $\ket{\up}$.  Given these time scales, we take  125 $\mu$s steps for piecewise constant evolution of the control waveforms.  

We compare the performance of the semi-analytical approach of Sec. III with the fully numerical search.  In all cases, we performed optimizations for 20 states chosen according to the Haar of measure on $SU(8)$ and then averaged the fidelity over the results for different states.   In the semi-analytical approach, we alternate $SU(2)$ rf- and microwave-generated rotations.  Each rotation is made robust  
by decomposing them into 3 steps of 125 $\mu$s duration whose amplitude and phases are found by  numerical optimization.  As the protocol does not have a specified end time, the number of rotations is not fixed.  For each state we repeat the optimization 10 times and pick the optimization that attains a chosen threshold fidelity in the shortest time.   In contrast, for the fully numerical approach, the microwave and rf fields are present simultaneously and we fix the total duration of the waveform, which sets the number of 125 $\mu$s steps.  We choose the duration to be long enough to achieve high fidelities; the duration is chosen for each case depending on the amount of inhomogeneity. 
We set the amplitude of the rf and microwave fields to their maximum values, and only optimize their phase. 
 The algorithm is then iterated until a target fidelity of 0.99 is reached.  

\begin{figure}
\includegraphics{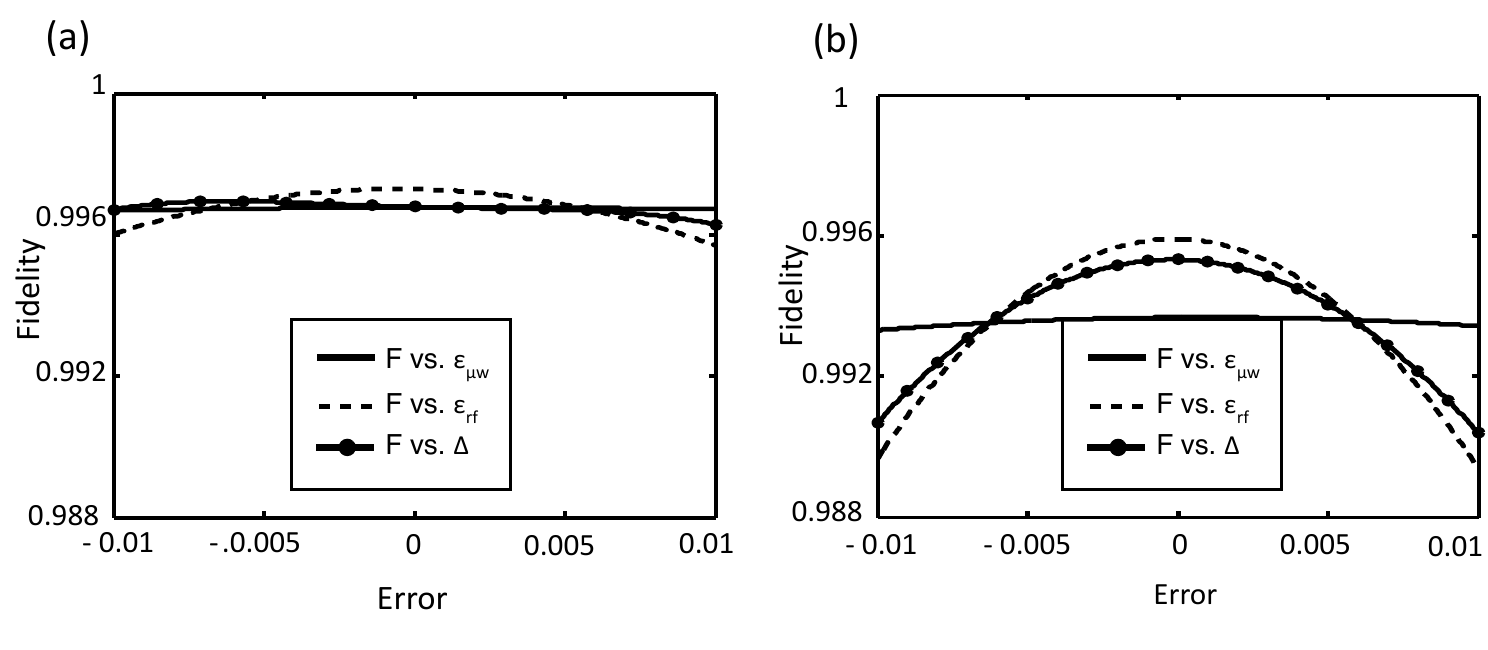}
\caption[rf and Microwave controls]{Average fidelity vs. inhomogeneity for (a) semi-analytic state preparation and (b) fully numerical state preparation.  The $x$-axis represents the fractional variation in the each parameter for the three cases -- power inhomogeneity in rf and micorwave, $\epsilon_{rf,\mu w}$, and detuning as a fraction of $\Omega_{\mu w}$. We show the performance as a function of one inhomogeneity and averaged over the the other two.  Each curve is an average of 20 random states in the 8 dimensional Hilbert space.  In the semi-analytic case, each waveform consists of a series of alternating rf and microwave $SU(2)$ pulses  for a total of 3.98 ms. The optimized pulses are found through a numerical search of the time varying amplitude and phase of the relevant fields.  In the fully numerical case, rf and microwave fields are applied simultaneously for a total of 1 ms.  The microwave and rf powers are set to their maximum values, and a we perform a numerical optimization of the phases.  For both the semi-analytic and fully numerical case,  the waveform consisted of a series of piecewise constant 125 $\mu$s pulses.}              
\label{Spread1PerCombinedBoth}
\end{figure}

A comparison of the results for errors of up to 1\% in  $\epsilon_{rf}$, $\epsilon_{\mu w}$ and $\Delta$ are shown in Fig.~\ref{Spread1PerCombinedBoth}.  In both the semi-analytical and fully numerical protocols, the pulse sequences are found by optimizing the fidelity on a coarse grid in parameter space defined by $\epsilon_{rf} = 0,\pm 0.01$, $\epsilon_{\mu w} = 0,\pm 0.01$ and $\Delta = 0,\pm 0.01\Omega_{rf}$, and then averaged.  To ensure that the pulse sequences perform as desired, we then calculate the fidelity on a finer grid with 15 evenly spaced points between $\pm 1\%$ in those parameters. For the semi-analytical case, the total time to reach a fidelity greater than 0.99 is 3.94 ms, averaged over the 20 states.  As can be seen from the figure, a fidelity of over 0.99 is maintained over the range of parameters, and the fidelity averaged over that range is 0.997.  For the fully numerical case, as seen in Fig.~\ref{Spread1PerCombinedBoth}, a fidelity of over 0.99 is maintained throughout the parameter range, and the fidelity averaged over that range is 0.994.  While this sample shows slightly lower average fidelity, the duration of the fully numerical waveform is only 1 ms long, nearly a factor of 4 speed up compared to the sequence of $SU(2)$ rotations.  Thus, we see that for inhomogeneities of 1\%, both approaches achieve high fidelity, though the sequence found via fully numerical optimization can do so in significantly less time.

\begin{figure}[!htp]
\includegraphics{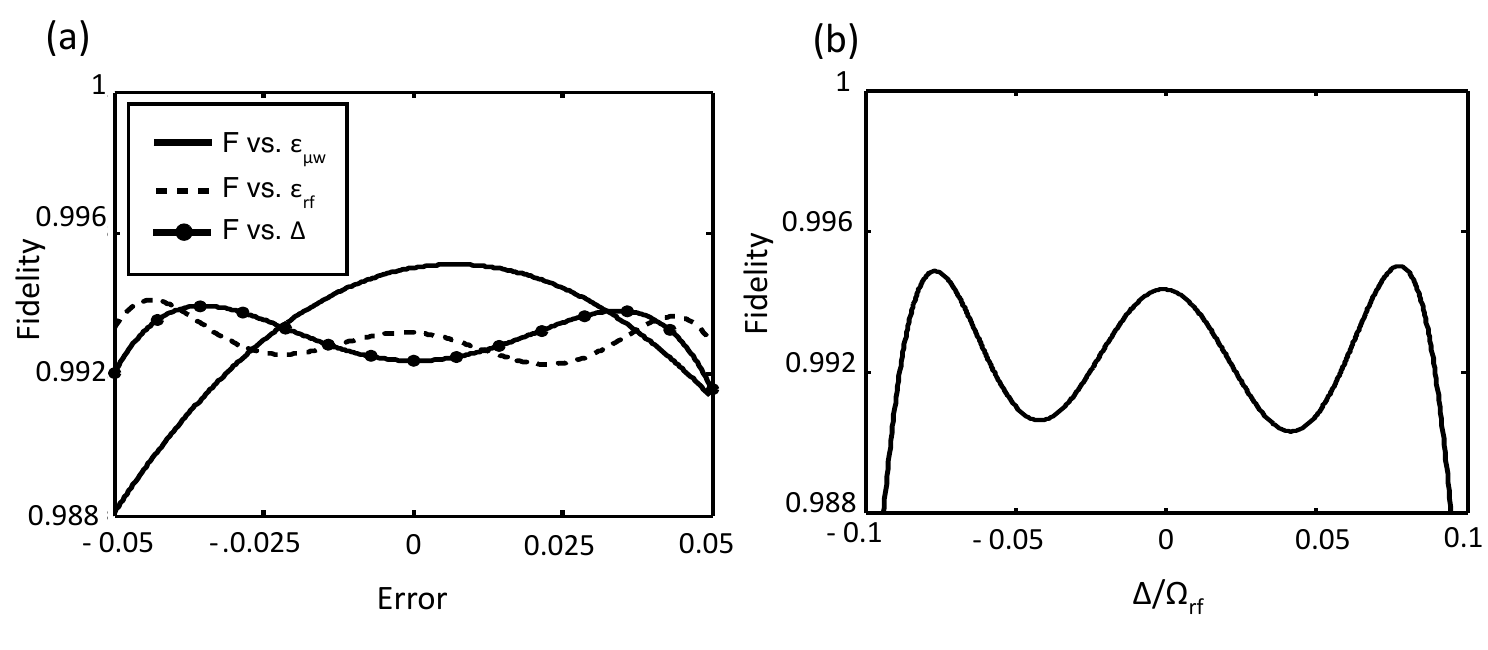}
\caption[Fidelity of Fully Numerical State Preparation vs 5\% Errors]{(a) Fidelity of fully numerical state preparation vs 5\% errors in $\epsilon_{rf}$, $\epsilon_{\mu w}$, and $\Delta/\Omega_{rf}$.    The $x$-axis represents the fractional variation in the each parameter for the three cases -- power inhomogeneity in rf and microwave, $\epsilon_{rf,\mu w}$, and detuning as a fraction of $\Omega_{\mu w}$. We show the performance as a function of one inhomogeneity and averaged over the the other two.  Each curve is an average of 20 random states in the 8 dimensional Hilbert space. (b) Fidelity of fully numerical state preparation vs 10\% errors in $\Delta/\Omega_{rf}$. }
\label{Spread5perCombinedSpread10per}
\end{figure}

While the semi-analytical method yield waveforms that require substantially more time to achieve high fidelity when compared with the fully-numerical method, it serves as an important proof-of-principle.  For the remainder of the paper we focus on the fully-numeric method.   Figure~(\ref{Spread5perCombinedSpread10per}.a) shows the results of fully numerical optimization with errors of 5$\%$.  In this case, the optimization was performed on a grid in parameter space defined by $\epsilon_{rf} = 0,\pm 0.05$, $\epsilon_{\mu w} = 0,\pm 0.05 $ and $\Delta = 0,\pm 0.05 \Omega_{rf}$.  The duration of the entire pulse sequence is 5~ms.  After the optimization, we calculate the fidelity on a finer grid of 15 points evenly spaced between $\pm 5\%$ for each of those errors, as shown in the figures.  A fidelity of over 0.988 is maintained throughout this range of parameters and we achieve an average fidelity of 0.993.  Optimization by the fully numerical method is sufficiently powerful to compensate errors well beyond experimental uncertainty.  As an example, in Fig.~(\ref{Spread5perCombinedSpread10per}.b) we present the results of optimizing errors of $\Delta$ that are  10\% of $\Omega_{rf}$.  We perform the optimization on a grid of 10 evenly space points between $\pm 0.1 \Omega_{rf}$.  The timing is the same as the previous example.  Fidelities of over 0.98 are maintained over the entire range of the detuning, while the fidelity averaged over the full range of errors is 0.992.   We summarize our results in Table~\ref{summary}.

\begin{table}
\begin{center}
\begin{tabular}{|c|c|c|c|}
\hline
Inhomogeneity Range & Optimization Method &  Total Time & Average Fidelity\\
\hline
$\epsilon_{rf},\epsilon_{\mu w},\Delta/\Omega_{rf}\in[-.01,.01]$   & Semi-Analytic  & 3.94 ms & 0.997\\
\hline
$\epsilon_{rf},\epsilon_{\mu w},\Delta/\Omega_{rf}\in[-.01,.01]$   & Fully Numerical  & 1.0 ms & 0.994\\
\hline
$\epsilon_{rf},\epsilon_{\mu w},\Delta/\Omega_{rf}\in[-.05,.05]$   & Fully Numerical   & 5.0 ms & 0.993\\
\hline
$\Delta/\Omega_{rf}\in[-.1,.1],\epsilon_{rf},\epsilon_{\mu w}=0$  & Fully Numerical   & 5.0 ms & 0.992\\
\hline
\end{tabular}
\caption[Table of published work.]{Summary of results.}
\label{summary}
\end{center}
\end{table}

\subsection{Tomographic adressing}
As discussed in the introduction, the tools of ensemble control allow us to design waveforms for tomographic addressing.  For example, a spatial gradient of the detuning imposed through an external field can be used to spatially address different regions of an ensemble. In our system, large gradients can be achieved by the fictitious magnetic field produced by the light shift associated with a circularly polarized laser beam \cite{deutsch_quantum_2009}.  A local unitary transformation can be designed as a function of the intensity of the addressing laser beam.

\begin{figure}[!htp]
\includegraphics{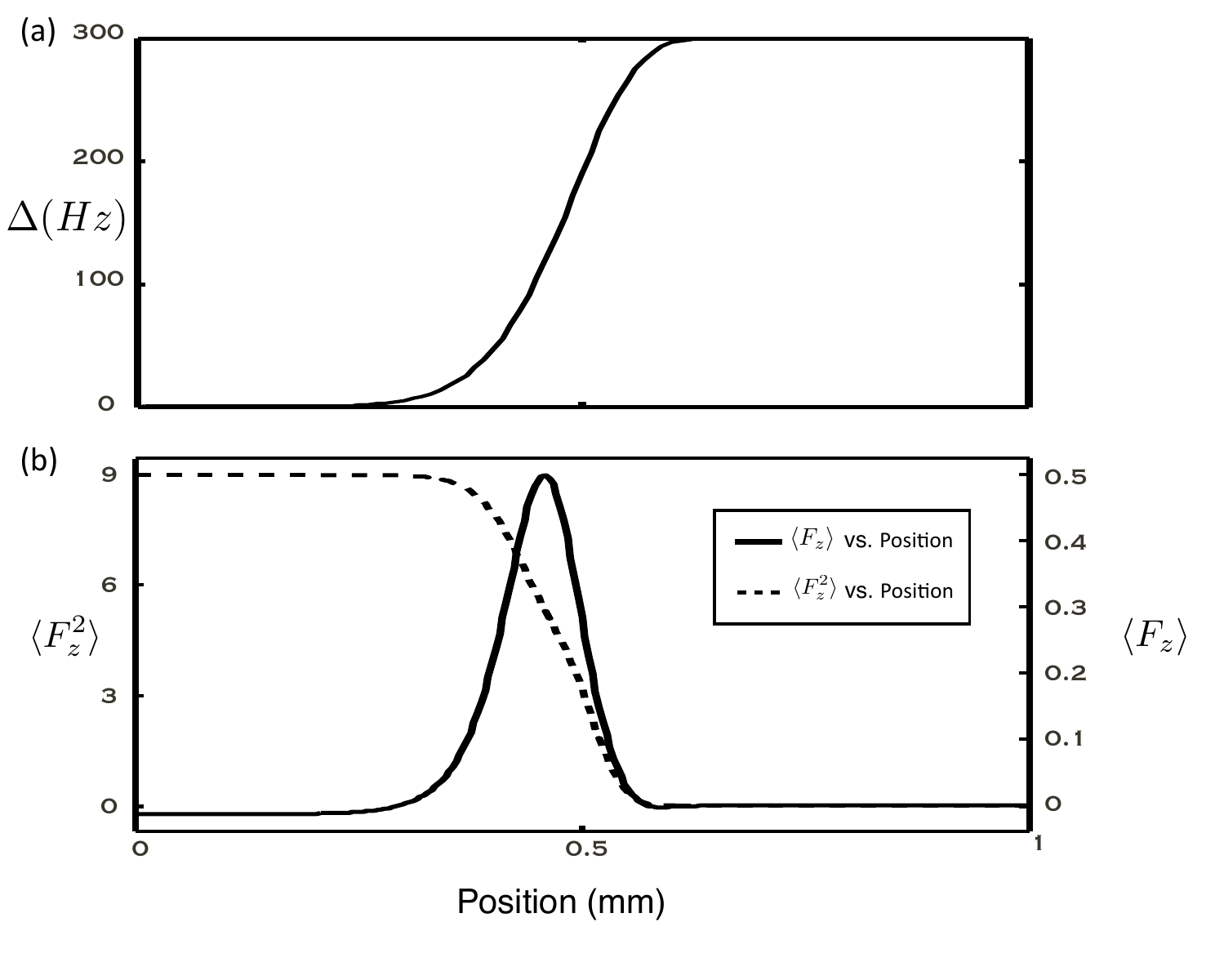}
\caption[Results of Spatially Selective Pulse]{Spatially selective control.  (a) Plot of the spatial variation in the detuning, $\Delta$,  imposed by a light-shift gradient.  The gradient separates the gas into two spatial regions in order to generate two distinct target states : for $x<0.5$ mm, the target state is  $\ket{\psi_1} = \frac{1}{\sqrt{2}}(\ket{F=3, m_F = -3} + \ket{F=3, m_F = 3})$ and for $x>0.5$ mm the target state is  $\ket{\psi_2} = \ket{F=3, m_F = 0}$.  Because a perfectly sharp transition between the two detunings is not possible in the lab, there is a transition region around $x=0.5$ mm.  The microwave and rf  fields are applied simultaneously with maximum power and numerically optimized phases.  (b) Results of the sythesized states.  The left axis shows $\langle F_z^2 \rangle$  vs. position and the right axis shows $\langle F_z \rangle$ vs. position.  If the target states are synthesized perfectly, then $\bra{\psi_1} F_z \ket{\psi_1}=\bra{\psi_2} F_z \ket{\psi_2}=0$  and $\bra{\psi_1} F_z^2 \ket{\psi_1} = 9$ while $\bra{\psi_2} F_z^2 \ket{\psi_2} = 0$.  With the exception of the transition region at $x=0.5$~mm, these goals are achieved.}
\label{spatial}
\end{figure}

As a proof-of-principle, we consider a cold atomic gas and prepare two different ``nonclassical" states (i.e. not spin-coherent states) in two spatial regions by applying global control pulses in the presence of a detuning inhomogeneity.  In this example we have chosen to synthesize $\ket{\psi_1} =\left(\ket{F=3,m_F=-3}+\ket{F=3,m_F=3}\right)/{\sqrt{2}}$ in one half of the gas and $\ket{\psi_2}=\ket{F=3,m_F = 0}$ in the other.  Such a distribution is highly nonclassical and in itself would be an interesting starting point for the study of dynamics in spinor gases \cite{sadler_2006}.  In the ideal case, we would apply a step function to perfectly select two regions with two distinct detunings through, e.g., application of a light-shift phase mask. In practice, however, the laser field which creates the level shifts cannot be focused to a perfectly sharp edge.  We take this into account by modeling the spatial variation in detuning by 
\begin{equation}
\Delta(x) = \Delta_0(1- e^{-(x/a)^m} ).
\end{equation}
In the limit $m \rightarrow \infty$ we recover a step function.  As an example we take $a=0.5$ mm and $m=8$,  which gives a transition region that is quite large compared to the wavelength of light and thus can be easily implemented with available optics.   As depicted in Fig.~(\ref{spatial}), we choose $\Delta_0 = 300$ Hz, which is large enough that the two regions can be easily distinguished, but is not large compared to the rf/microwave Lamor/Rabi frequencies.   

Our goal is to synthesize $\ket{\psi_1}$ in the region with detuning $\Delta_0=0$ and $\ket{\psi_2}$ in the region with $\Delta_0=300$ Hz.
We have included some spread in the detuning to account for possible noisy background fields, and optimize on a grid around the desired detuning, $\Delta=\Delta_0+\delta$, $\delta =\{0,\pm10 \rm{Hz}\}$.  In this case the objective function is,
\begin{equation}
\mathcal{F}( \mathbf{\Phi})=\sum_{\delta} \left( |\bra{\psi_1 }U(\Delta=\delta,\mathbf{\Phi})\ket{\up}|^2+  |\bra{\psi_2}U(\Delta=300\text{Hz}+\delta,\mathbf{\Phi})\ket{\up}|^2 \right).
\end{equation}
In the transition region between the two targets, we do not constrain the solution.  Using the same gradient search, we find the control waveform for the phases of the oscillating fields with their amplitudes fixed at the maximum values.

The results of the optimization are shown in Fig.~(\ref{spatial}).  To evaluate the performance we have plotted both $\langle F_z \rangle$ and $\langle F_z^2 \rangle$ as a function of position.  For the two target states we have $\bra{\psi_1} F_z \ket{\psi_1}=\bra{\psi_2} F_z \ket{\psi_2}=0$  and $\bra{\psi_1} F_z^2 \ket{\psi_1} = 9$ while $\bra{\psi_2} F_z^2 \ket{\psi_2} = 0$.  In the region between $x=$ 0 and 0.3 mm, where the detuning is between $\Delta/2\pi=$ 0 and  10 Hz, $|\langle F_z \rangle| \leq 0.0125$ and $\langle F_z^2 \rangle \geq8.95$ as expected for $\ket{\psi_1}$.  In the region between $x=$ 0.6 and 1 mm, where the detuning is between $\Delta/2\pi=$ 290 and 300 Hz, $|\langle F_z \rangle| \leq 1.5 \times10^{-3}$  and $\langle F_z^2 \rangle \leq0.012$, as expected for $\ket{\psi_2}$.  

\section{Summary \label{sec:summary}}
We have studied the control of $d$-level quantum systems or qudits encoded in the hyperfine magnetic sublevels of alkali-metal atoms when there are variations in the external rf and microwave fields that drive the system.  Such variations could be the result of uncertainties that result from experimental imperfection or intentionally applied in order to tomographically address specific members of an inhomogeneous ensemble. We restricted our attention to qudits formed from the subspace of the lower hyperfine manifold $F_{\dn}=I-1/2$, plus one sublevel in the upper manifold $F_{\up} = I+1/2$; for $^{133}$Cs this is an 8D Hilbert space.  With this structure, we can achieve control on the system through a simple series of $SU(2)$ rotations driven alternately by resonant rf and microwave fields.  

We studied the simplest control problem  -- state-synthesis -- whereby a known fiducial state is transformed to an arbitrary state in the Hilbert space.  Our construction is based on the protocol of Law an Eberly \cite{law_arbitrary_1996}, extended to $F>1/2$ representations of $SU(2)$.   Moreover, because it consists of a series of $SU(2)$ we can draw upon the results of \cite{kobzar_pattern_2005, li_control_2006}, which proved that arbitrary unitary transformations on two-level systems could be constructed as a function of variations in the detuning and Rabi frequency of the driving Hamiltonian, and we extended them to our case of inhomogeneous control of qudits.

We proved that the semi-analytic protocol is efficient, though not necessarily optimal in the duration of the pulse sequence.  More importantly, this procedure provides a foundation for numerical searches for more efficient pulse sequences.   This situation is similar to that found in two-level control, where a Lie algebraic approach provides a proof of principle that appropriate controls exist, while numerical optimization is used to find the optimum pulse sequence \cite{kobzar_pattern_2005, li_control_2006}.  The search is based on a simple gradient ascent with the objective of maximizing the fidelity of the target, averaged over the inhomogeneous parameters.  We compared the performance of semi-analytical state synthesis to fully numerical state synthesis, and studied robust control over a range of experimental uncertainties.  For small inhomogeneities, $< 1\%$,  both approaches achieve average fidelities greater than 0.99.  However, the fully numerical approach can find control sequences which require significantly less time. Thus, for the remainder of the paper, we studied the fully numerical approach. For the parameters we studied, fidelities greater than 0.99 are possible with $5\%$ errors in detuning and amplitudes of the driving fields.  As a testament to the efficiency of the search procedure, we were able to find control waveforms that are robust to $10\%$ errors in detuning, well beyond experimental uncertainty.

In addition to robust control we performed a proof-of-principle test of tomographic addressing using designed spatial variations in the detuning over an extended ensemble.  Such tomography has been employed to address individual atoms in optical lattices, and considered even under circumstances beyond the diffraction limit of an addressing laser beam \cite{weitenberg_2011}.  Prior work assumed that the non-addressed members were unaffected solely because they were too far off resonance.  Here we studied more general extensions, employing the tools of ensemble control.  In practice, different members of the ensemble can be made to undergo different unitary transformations depending on the local parameters.  As a simple example, we showed how one can synthesize, with high fidelity, the state $\ket{\psi_1} = (\ket{F=3,m_F=-3}+\ket{F=3,m_F=3})/\sqrt{2}$ in one half of the gas and $\ket{\psi_2}=\ket{F=3,m_F = 0}$ in the other, limited by the  resolution of the spatial gradient of detuning.

In the future, we intend to extend this work in a number of directions.  We have previously shown how one can leverage off of the efficiency of state-synthesis to design the a full unitary transformation of the $d$-dimensional Hilbert space \cite{merkel_constructing_2009}. For such a procedure to work, it will be critical take into account inhomogeneities in the driving fields.  The tools of ensemble control developed here provide the necessary foundation.  Additionally, we plan to extend the work of tomographic control towards addressable control of atoms in optical lattices, a key ingredient in many studies of quantum computing and quantum simulation.

\emph{Acknowledgments}.  We thank Poul Jessen, Aaron Smith and  Brian Anderson for helpful discussions.  BEM was supported NSF Grant NSF Grant PHY-0903692, the EU integrated project AQUTE, and the IARPA MQCO program.  IHD was supported by NSF Grant PHY-0903692 and the CQuIC NSF grant PHY- 0903953.  STM was supported by NSERC through the discovery grants and QuantumWorks. This research was also funded by the Office of the Director of National Intelligence (ODNI), Intelligence Advanced Research Projects Activity (IARPA), through the Army Research Office.


\bibliographystyle{apsrev4-1_no_url_isbn_issn_3}
\bibliography{qudit_v4}

\end{document}